\pdfoutput=1

\documentclass[10pt,conference,a4paper]{IEEEtran}
\usepackage{multicol}
\usepackage{hyperref}
\usepackage{url}
\usepackage{stfloats}
\usepackage{graphicx}
\usepackage{subfigure}
\usepackage{amsfonts}
\usepackage{bbm}
\usepackage{url}
\usepackage{amsmath}
\usepackage{cases}
\usepackage{multirow}
\usepackage{diagbox}

\usepackage[noend]{algpseudocode}
\usepackage[ruled,vlined, linesnumbered]{algorithm2e}
\usepackage{tikz}
\usepackage{booktabs}
\usetikzlibrary{positioning}
\usepackage{ragged2e}
\usepackage{color,soul}
\usepackage{array}
\usepackage{chngpage}
\usepackage{multicol}
\usepackage{caption}
\usepackage{enumitem}
\usepackage{amsmath}

\newtheorem{definition}{Definition}

\DeclareMathOperator*{\argmax}{arg\,max}

\DeclareMathOperator*{\onehot}{OneHot}
\DeclareMathOperator*{\minimize}{minimize}

\ifCLASSINFOpdf
\else
\fi
\begin{document}
%
\title{Conditional-UNet: A Condition-aware Deep Model for Coherent Human Activity Recognition From Wearables}

\author{\IEEEauthorblockN{Liming Zhang}
\IEEEauthorblockA{College of Science\\
George Mason University\\
Fairfax, Virginia 22030\\
Email: lzhang22@gmu.edu}
}


%


\maketitle
\pagestyle{plain}

\begin{abstract}
Recognizing human activities from multi-channel time series data collected from wearable sensors is ever more practical. However, in real-world conditions, coherent activities and body movements could happen at the same time, like moving head during walking or sitting. A new problem, so-called ``Coherent Human Activity Recognition (Co-HAR)'', is more complicated than normal multi-class classification tasks since signals of different movements are mixed and interfered with each other. On the other side, we consider such Co-HAR as a dense labelling problem that classify each sample on a time step with a label to provide high-fidelity and duration-varied support to applications. In this paper, a novel condition-aware deep architecture ``Conditional-UNet'' is developed to allow dense labeling for Co-HAR problem. We also contribute a first-of-its-kind Co-HAR dataset for head movement recognition under walk or sit condition for future research. Experiments on head gesture recognition show that our model achieve overall $2-3\%$ performance gain of F1 score over existing state-of-the-art deep methods, and more importantly, systematic and comprehensive improvements on real head gesture classes.
\end{abstract}


%
\IEEEpeerreviewmaketitle

\section{Introduction}
\begin{figure}
    \centering
    \includegraphics[width=\linewidth, trim={2cm 7cm 4cm 7cm},clip]{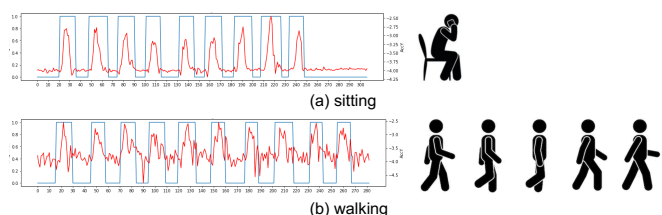}
    \caption{A toy example of coherent human activity recognition and signals (blue lines are boolean head gesture labels, red lines are one accelerometer data): a) performing a gesture under sitting; b) performing a gesture under walking.}
    \label{fig:example}
\end{figure}

With the rapid development and lower cost, wearable devices with embedded sensors are becoming more and more popular to be used for a plethora of applications, such as healthcare~\cite{gray2007head}, authentication~\cite{parimi2018analysis}, robotic control~\cite{gray2007head}, virtual/augmented reality~\cite{zolkefly2018head,hachaj2019head,chen2019augmented} and e-learning~\cite{deshmukh2018feedback}. 
Although there are many possible sensors in many real-world applications of wearables, people prefer to have limited devices with multiple functions, such as smart phone, virtual reality headset, smart glass, or wireless headphone, instead of wearing multiple devices at the same time. Another challenge is that body moves simultaneously during daily activities that generate complicated mixed signals for those limited devices mounted on the body. These multiple human activity and movements are interactively interfered with each other. For example, such a challenging task is to recognize head gestures during walk or sit conditions (Figure \ref{fig:example}) using embedded accelerometer and gyroscope sensors in only one location, for example a headphone or Virtual Reality headset. The denoised signal collected under sitting situation (above red line) has a clearer pattern compared to signal under walking situation (the lower red line). This is a relatively hard task, because body parts except head during walking generates stronger inertia than the simultaneous head movements. Many previous works on head gestures \cite{gray2007head, wu2017applying, hachaj2019head} only focus on a controlled sitting situation, or they simply do not consider such coherent activities at all. In this work, a new problem, so-called `` Coherent Human Activity Recognition'' (Co-HAR), is firstly proposed for such kind of tasks with interference movements.

Recent development of deep learning shed a light on human activity recognition research, since deep learning allows to learn latent features using deep structures such as convolution layer, pooling layer and embedding layer~\cite{goodfellow2016deep}. It usually requires much less or even no effort on feature extraction than pre-dated models before deep learning. In an end-to-end fashion, deep learning models have better generality which allow a model structure to be performed well in different data without domain-specific work that resulted in shorter development cycles. Deep learning could also be naively applied for Co-HAR. However, deep learning for Co-HAR problem has not been systematically explored and specially-designed.

Beyond naively using deep methods, some critical technical challenges prevent current deep architectures to perform better, including: \textit{1) single location of sensors has mutual impact of signals.} As discussed, sensors are placed only in a headphone over a user's head. It is impractical to ask a user to wear sensors all over body in the real-world scenario. It is also difficulty to exactly separating signals and reduce mutual impacts using existing approaches like a basic multi-label classification \cite{vaizman2018context,mohamed2020multi}; \textit{2) imbalanced domination of different activities could fade away signals of other activities.} Sensors could have different sensitive levels to different body movements. The dominating movement might not be the most interesting one that we would like to examine. For example, in the head gesture problem (in Figure \ref{fig:example}), sensors are placed in a headphone, walking generates strong signals in the forward inertia. However, head gestures are more critical for most applications like Virtual Reality. The current models might have limited power in such scenarios. \textit{3) multi-label window problem for duration-varied activities}. The time steps in one window may not always share the same ground truth label, and the duration of an activity always vary in different windows. Mixing of ground truth labels not only creates difficulty for underlying models but also reduces flexibility of usages due to a whole set of hyper-parameters to be considered, such as the best window length, sampling stride and window labelling strategy.

By tackling these challenges for Co-HAR problem, we proposed a novel condition-aware deep structure, called ``Conditional-UNet'', which could take multi-channel sensor data embedded only at one location of human body as input and hence explicitly capture conditional dependence within coherent labels. We propose a novel encoding module to model the conditional dependence which could reduce the mutual impact of coherent body movement and guide our model to learn better patterns in activities with imbalanced domination. Since it follows the dense labeling approach \cite{yao2018efficient}, it avoids multi-label window problem, but it aims to classify multiple labels which is more challenging than previous dense labeling works. The major contributions of our work are summarized as follows:
\begin{itemize}[nolistsep,leftmargin=*]
\item We consider a challenging problem, so-called ``Coherent Human Activity Recognition'' (Co-HAR) which classify coherent activity labels beyond simple multi-label scenarios and use multi-channel time-series data from one wearable device only at one location of human body.

\item A novel condition-aware deep classification model, called ``Conditional-UNet'', is developed to better densely classify coherent labels. Conditional-UNet explicitly models conditional dependence through a novel deep structure, including a new encoding module with specially-designed gradient-permitted sampling and embedding structure and a UNet-based decoding module.

\item The contribution of a new dataset for Co-HAR problem. To conduct experiments, we build an Ardino-based device to collect data and label head gestures and walk/sit condition.

\item Extensive experiments show that our proposed Conditional-UNet outperform existing state-of-the-art UNet model by $2-3\%$ of F1 score over head gesture label classification.
\end{itemize}

\section{Related Work}
\textbf{Feature Extraction based Methods.} 
Pre-dated models before deep learning rely heavily on hand-crafted features (e.g., mean, variance, kurtosis, or other kinds of indexes) \cite{wu2017applying,parimi2018analysis}, motion (e.g., physical laws) \cite{zolkefly2018head}, transform-based feature (e.g., wavelet \cite{chung2009realtime}, fourier transform \cite{gamal2013hand}). Exacted features are then feed to classifiers such as Support Vector Machines \cite{zolkefly2018head}, Boosting Tree \cite{wu2017applying} and Hidden Markov Model \cite{zolkefly2018head}. These approaches usually work well for a specific type of tasks while fails for other types of applications. 

\textbf{Deep Learning based Dense Labeling.} 
With the advancement of deep learning methods, the applications of deep learning to HAR using data from wearable sensors are relatively new. More and more works propose to utilize some kinds of deep learning methods \cite{yao2018efficient,ignatov2018real,ferrari2019human}. The success of deep-learning-based methods comes from their high expressiveness in learning underlying complex principles directly from the data in end-to-end fashion without handcrafted rules. Another most recent advancement by using deep learning is the Dense Labeling \cite{yao2018efficient} using a fully convolutional network \cite{long2015fully} to label each sample instead of a sliding window. It avoids the segmentation problem in most of conventional methods. Another work \cite{zhang2019human} achieves the same goal of dense labeling but utilizes another deeper structure called ``UNet''. These works still assumes a single activity label rather than coherent activities. 

\textbf{Multi-label classification}
A recent work \cite{varamin2018deep} for multiple overlapping labels output activity sets of multiple labels for each window using deep neural networks, but learned labels are not associated with each sample and there is not explicitly consideration of conditional dependence in different human activities. Another recent work \cite{mohamed2020multi} converts multiple labels into one label with all classes from different labels to solve the multi-label classification. There is no existing works considering coherent multiple labels which have conditional dependence within them.
We point readers to more details in other survey papers in this area \cite{nweke2018deep, chen2020deep}.

In summary, deep learning methods, including the state-of-the-art UNet, are actively researched in exiting works with better performances than pre-dated conventional methods. However, to our best knowledge, there are no work considering  conditional dependency in multiple dense labels beyond simple multi-label classification. Next, we would formally define our problem.

\section{Problem Definition}
A set of sequences $\mathcal{D} = \{ (\pmb{X}^{(i)}, \pmb{Y}^{(i)}) \}_i^N, \forall \pmb{X}^{(i)} \in \mathbb{R}^{K \times T^{(i)}}, \pmb{Y}^{(i)} \in \mathbb{R}^{H \times T^{(i)}}$, which contains a multivariate sequence $\pmb{X}^{(i)}$ which have $K$ variables of sensors and the time length of each sequence is $T^{(i)}$. Here, each time step $t \in \{1, \dots, T^{(i)}\}$ is normally referred as a sample. Correspondingly, $\pmb{Y}^{(i)}$ is a multi-label sequence with $H$ labels. For each label $h \in \{1,\dots, H\}$, there are $C_h$ different numbers of classes for this label $h$, so an element $\pmb{Y}^{(i)}_{h, t} \in \{1, \dots, C_h\}$, where $\pmb{Y}^{(i)}_{h, t} = 1$ usually define for a null label (e.g. no hand gesture is performed). The sequence index $(i)$ is dropped in later parts whenever it is clear that we are referring to terms associated with a single sequence. We define our problem as follow:
\begin{definition}[Coherent Human Activity Recognition]\label{def:co-har} is a multi-label classification problem with conditional dependency assumption within joint multiple coherent labels and has a goal to minimize the difference between a classifier's predicted $H$-label sequence $\pmb{\hat{Y}}^{(i)}$ for $K$-channel sample sequence $\pmb{X}^{(i)}$ and the ground truth label $\pmb{Y}^{(i)}$ given multi-channel time-series sequences set $\mathcal{D}$.
\end{definition}
\begin{figure}
    \centering
    \includegraphics[width=\linewidth, trim={0cm 11cm 5cm 0cm},clip]{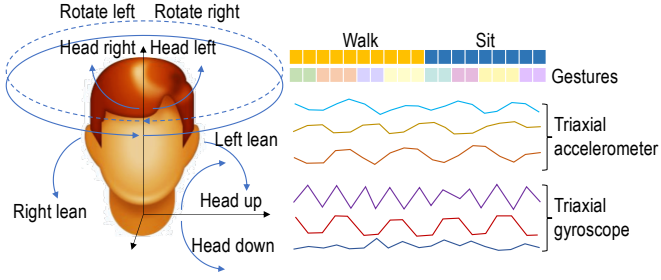}
    \caption{A example task of Co-HAR: recognizing head gesture using accelerometer and gyroscope sensors on headphone. It has two coherent labels: head gesture label and walk/sit label.}
    \label{fig:problem}
\end{figure}

For example, in our head gesture task, a sample $\pmb{X}^{(i)}_t$ contain $K = 6$ variables of sensors including tri-axial acceleration and tri-axial gyroscopes. If a sampling rate is $1~Hz$, then there should be $T^{(i)} = 60$ total measures for a one-second window. Since we are interested in two types of labels, head gesture and walk/sit condition, so there are two label types $L = 2$. For walk label $y_1 \in \{1, 2\}$, $y_1 = 1$ if a human subject is sitting, and $y_0 = 2$ indicates a walk activity. Similarly, for head gesture label $y_2$, $y_2 = \{1, \dots 9\}$, and $y_2 = 1$ means no head gestures, and other numbers indicates other $8$ head gestures, and $C_1 = 9$. 

A key component of Co-HAR is that we keep the complete joint probability of all labels $p(Y_1^{(i)}, \dots, Y_H^{(i)})$ with conditional dependency, not simplified joint probability $p(Y_1^{(i)}) \times \dots \times p(Y_H^{(i)})$ that assumed independence of all labels in the basic multi-label classification. More details are in following methodology Section \ref{sec:framework}.

\section{Methodology}
In this section, condition-aware framework of Co-HAR and detailed modeling components of Conditional-UNet are introduced. Section~\ref{sec:framework} shows the condition-aware deep framework for Co-HAR problem and the formalization of its decomposed conditional losses. 
In Sections \ref{sec:sampling}, \ref{sec:embedding}, and \ref{sec:merging}, three key components of handle joint label conditions are introduced to capture the conditional dependence within coherent activity labels. 

\subsection{Condition-aware deep framework}
\label{sec:framework}
In this part, condition-aware deep framework is developed firstly to build up a probabilistic understanding of Co-HAR problem, and to formalize loss with an independent loss and a series of dependent losses. 
In general, the goal of Co-HAR in previous Co-HAR definition \ref{def:co-har} is to learn a joint probability of multiple labels given multi-channel sensor data, noted as $p(Y_1, \dots, Y_H | X)$. By an axiom of probability, joint probability of observing a sequence can be decomposed to a series of independence and dependence components in Equation \ref{eq:joint_probability}: 
\begin{equation}
\begin{split}
&p(Y_1, \dots, Y_H | X) = \\
&p_{\theta_1}(Y_1 | X) p_{\theta_2}(Y_2| Y_1, X) \dots p_{\theta_H}(Y_H | P_{H-1}, \dots, Y_1, X)
\end{split}
\label{eq:joint_probability}
\end{equation}
where $\theta_i$ are parameters of each probability function $p(\cdot)$. This is the complete conditional relationship for Co-HAR. By assuming conditional independence between all the labels, we can simplify it to be $p(Y_1, \dots, Y_H | X) = p_{\theta_1}(Y_1 | X) p_{\theta_2}(Y_2| X) \dots p_{\theta_H}(Y_H | X)$, which is a normal multi-label classification framework in many current works \cite{mohamed2020multi,vaizman2018context}. However, in this work, we want to keep the conditional dependence since conditional independence assumption drop a lot of useful information. Next, the condition-aware loss function is introduced with its different components.

\begin{figure}
    \centering
    \includegraphics[width=\linewidth, trim={0cm 7.5cm 6cm 0cm},clip]{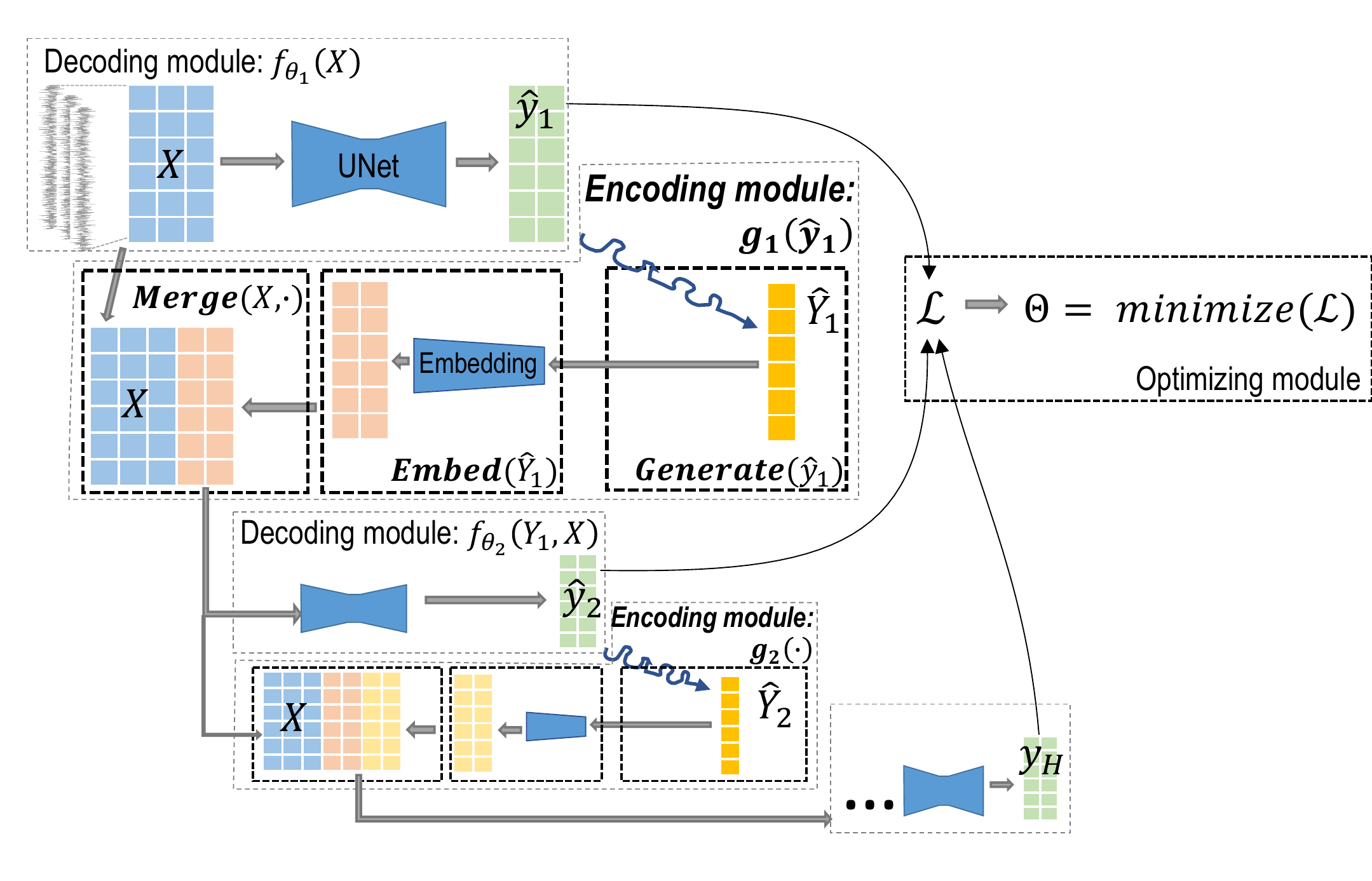}
    \caption{Conditional-UNet: a conditional deep model with UNet module, Sampling module, and embedding module to capture conditional dependence in coherent activities.}
    \label{fig:deep_model}
\end{figure}

\textbf{Condition-aware multi-label dense classification loss:} following the common approach of Maximum Likelihood Estimation (MLE) \cite{bishop2006pattern} and similar to dense labeling \cite{yao2018efficient}, we get loss function by factorizing joint probability in Equation \ref{eq:joint_probability} to each temporal sample with each label, and get its logarithmic transformation as follows:
\begin{equation}
\begin{split}
\mathcal{L} = &log(p(Y_1, \dots, Y_H | X)) = 
\sum_t^T \big( log(p_{\theta_1}(Y_{1,t} | X)) + \\
& 
\dots + log(p_{\theta_H}(Y_{H,t} | P_{H-1, t}, \dots, Y_{1,t}, X))\big)
\end{split}
\label{eq:log_probability}
\end{equation}
where $log(p_{\theta_1}(Y_{1,t} | X))$ is log-likelihood to observe different classes of label $1$ on time step $t$th sample. Furthermore, this log-likelihood with each class $m$ of a label is formulated as $\sum^{C_1}_m y_{1,t}^{m} log(p_{\theta_1}(Y_{1, t} = m | X))$, where $y_{1,t}^m$ is observed frequency of class $m$ in all samples, and $p_{\theta_1}(Y_{1, t} = m | X)$ is estimated likelihood of class $m$ got from deep model. The calculation of estimated likelihood is the same as deep model take sensor data as input and output estimated likelihood, noted as $\pmb{\hat{y}}_1 = f_{\theta_1}(X)$, where $\pmb{\hat{y}}_1$ is estimated logit vector with $m$th element on $t$th sample $\hat{y}_{1,t}^m$ as estimated probability of multi-label categorical distribution of label $1$. Until now, it is a normal MLE with more details in \cite{bishop2006pattern}. The difference of our condition-aware model start from label $2$. Instead of only taking $X$ as inputs for $Y_2$, our deep model takes conditional signals of $Y_1$ as input too, noted as $\pmb{\hat{y}}_2 = f_{\theta_2}(X, Y_1)$, or $\pmb{\hat{y}}_i = f_{\theta_i}(X, Y_1, \dots, Y_{i-1})$. It means that our condition-aware deep model should decode all previous labels as joint conditional dependence for the next label estimation. Before we show how conditional dependence is estimated, we summarize our condition-aware loss as follows:
\begin{equation}
\begin{split}
    &\mathcal{L} = - \frac{1}{N} \big( 
    \sum^T_t \sum^{C_1}_m y_{1,t}^m log(\hat{y}_{1,t}^m) \\
    &+ \sum^T_t \sum^{C_2}_m y_{2,t}^m log(\hat{y}_{2,t}^m) + \dots + \sum^T_t \sum^{C_l}_m y_{H,t}^m log(\hat{y}_{H,t}^m)
    \big)
    \\
    &\hat{y}_{1,t}^m = f_{\theta_1}(X), \hat{y}_{2,t}^m = f_{\theta_2}(Y_1, X), \\
    &\dots,  \hat{y}_{H,t}^m = f_{\theta_H}(Y_1,\dots,Y_{H-1}, X)
\end{split}
\label{eq:losses}
\end{equation}
\begin{equation}
\minimize_{\Theta} {\mathcal{L}}
\label{eq:minimize}
\end{equation}
where $\Theta = \{\theta_1, \dots, \theta_H\}$ is the set of all deep model's parameters that minimize negative log-likelihood loss $\mathcal{L}$ in Equation \ref{eq:losses}. 
\textbf{To model the joint label conditions within this condition-aware deep model, we create a chain of conditional deep models $f_{\theta_i}$, except the first $f_{\theta_1}$.} It follows the procedure illustrated in Figure~\ref{fig:deep_model}: 

\textbf{1) Decoding module:}
uses a deep encoding module $f_{\theta_1}(X)$ to compute the logit vector of label $1$ for each sample $\pmb{y}_{1} = f_{\theta_1}(X)$. Here, the deep encoding module is UNet \cite{ronneberger2015u}. UNet is originally designed for image segmentation following the idea of fully convolutional network \cite{long2015fully}. It is a deep fully convolutional network, which internally contains multiple down-sampling convolutional layers and multiple up-sampling deconvolutional layers. It is recently used to boost performance in normal HAR task \cite{zhang2019human} as a more powerful alternative than basic fully convolutional network using in Dense Labeling \cite{yao2018efficient}, CNN \cite{xu2018human}, or SVM \cite{zhang2019human}. We utilize the same structure as \cite{zhang2019human} with more details; 

\textbf{2) Encoding module:}
uses a decoding module to convert logit vector $\pmb{y}_{1}$ to a conditional signal. It includes three sub-modules in this module: a) a generating sub-module $Generate(\cdot)$ to get a sampled class for the first label from the categorical distribution, $\hat{Y_1} = Generate(\pmb{y}_1)$. It is detailed in Section \ref{sec:sampling}; b) an embedding sub-module $Embed(\cdot)$ (detailed in Section \ref{sec:embedding}) is used to project $Y_1$ to a continuous embedding space with $Embed_{\phi_1}(\hat{Y_1})$, where $\phi_i$ is the set of parameters; c) a merging sub-module $Merge(\cdot)$ (detailed in Section \ref{sec:merging}) to merge both $X$ and embedded signal as input into the next encoding module to get logit vector $\pmb{y}_{2}$ of label $Y_2$, $\pmb{y}_{2} = f_{\theta_2}(Merge(X, Embed(Cat(\pmb{y}_1))))$. For $Y_3$, the only difference is that it merges both $g_{\phi_1}(\hat{Y_1})$ and $g_{\phi_2}(\hat{Y_2})$ with $X$, $\pmb{y}_{3}$ of label $Y_3$, $\pmb{y}_{3} = f_{\theta_3}(Merge(X, Embed(Cat(\pmb{y}_1))), Embed(Cat(\pmb{y}_2))))$. we continue this chain of processes until it reaches the last conditional model for the last label $Y_H$. Here, all labels $Y_i$ are one-hot vectors;

\textbf{3) Optimizing module:}
uses all logit vector $\hat{\pmb{y}}_i$ to calculate the multi-label dense classification loss $\mathcal{L}$, and minimize it through gradient back-propagation optimization techniques for deep neural network models, such as Adam \cite{goodfellow2016deep}, or Stochastic-Gradient-Descent \cite{goodfellow2016deep}. Since we develop our code using PyTorch \cite{ketkar2017introduction}, the Adam Optimizer is directly used to perform optimization and train our Conditional-UNet classification model.

Module 1) and 3) are conventional works with more details in other works \cite{ronneberger2015u,zhang2019human,goodfellow2016deep}, while Module 2) is our main structure to explicitly handle conditional dependence. We will now introduce different parts of this novel decoding module.

\subsection{Gradient-permitted generating sub-module}
\label{sec:sampling}
This generating sub-module, noted as $Generate(\cdot)$ in Figure \ref{fig:deep_model}, is the first step to incorporate conditional dependence information in Conditional-UNet. Its goal is to generate a sample of current label $\hat{Y}_i$ from estimated logit $\hat{y}_i$ of which each element is a probability to get $m$th class, so that sampled label can be used as a conditional input for the next decoding module. The keys here are both to allow gradient back-propagation and to better process conditional dependence signal. We proposed two variants for this sub-module as follows:

\subsubsection{\textbf{Naive-Max trick}} which selects the maximum probability class $m$ in estimated logit vector $\hat{y}_i$ in Equation \ref{eq:naive_max}:
\begin{equation}
\begin{split}
    \hat{Y}_i &= \argmax_m \hat{y}_i, \\
    &\forall m \in \{1, \dots, C_h\}, \hat{y}_i \in \mathbb{R}^{C_h}
\end{split}
\label{eq:naive_max}
\end{equation}
where $C_h$ is the number of classes in label $Y_i$.
This Naive-Max trick simplifies a categorical distribution to focus only on its class with maximum probability, however, it does not capture the whole distribution information. For example in Figure \ref{fig:gumbel_max} (a), it can potentially learn a flatten distribution (differ from the true distribution in Figure \ref{fig:gumbel_max} (c)), whose class with the max probability does not differ a lot from other classes. In this case, the distinguishing power could be vanished because of the big variance in this approach. The Naive-Max does not block gradient flow, however, it is potentially unstable because the maximum class shift a lot during training process. If a maximum-probability class is changed in a followed training iteration, the gradient flowing path changes to the other class which has maximum probability in that training iteration. This is a huge instability disadvantage, while its advantage is its easy implementation.

\subsubsection{\textbf{ Gumbel-Max trick}} 
\begin{figure}
    \centering
    \includegraphics[width=\linewidth, trim={0cm 9cm 8.5cm 0cm},clip]{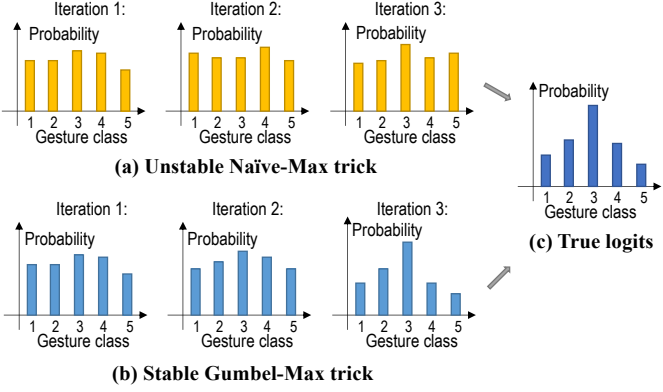}
    \caption{Stability illustration of different Gradient-permitted generating sub-module variants: (a) three iterations of Naive-Max trick, whose peak might bounce back and forth with a flatten distribution; (b) three iterations of Gumbel-Max trick, whose learn the whole shape of distributions with more stable process; (c) the true logits of categorical distribution to be learned.}
    \label{fig:gumbel_max}
\end{figure}

which implements the true process of sampling a class $\hat{Y}_i$ from a categorical distribution using logit vector $\hat{y}_i$, noted as $\hat{Y}_{i} \sim Cat(\hat{y}_i)$. This sampling process captures the full distribution information because it can still generate classes which do not have the maximum probabilities. In this way, we approximate the full categorical distribution for each class, not only the one with maximum probability. Unfortunately, this operation of sampling from categorical distribution do not have gradients, so it prevents gradients flowing in back-propagation training. The work-around solution is to use re-parameterization trick, specifically we leverage a ``Gumbel-Max'' trick \cite{jang2016categorical} for categorical data of labels. For example in Figure \ref{fig:gumbel_max} (b), Gumbel-Max reduces the chance of flatten distribution like Naive-Max, while pushes the distribution to concentrate on one or a few classes and decrease the probabilities of other classes. It can get a distribution closer to the shape of true distribution (Figure \ref{fig:gumbel_max} (c)), not only capture the peak one.

The Gumbel-Max trick is done in this way.
Specifically, we create $\pmb{y}_{i, t}' = \tanh((\pmb{q}^{i,t} + \pmb{g}) / \tau)$, where $\tau$ is so-called ``temperature'' hyper-parameter. Each value $g_j$ in $\pmb{g}$ is an independent and identically distributed (i.i.d.) sample from standard Gumbel distribution \cite{jang2016categorical}. $\pmb{g}$ has the same dimension as $\pmb{y}_{i, t}$.
We generate a one-hot representation $\pmb{y}_{i, t}$, whose $j$th element is one and all the others are zeros, where $j$ is got as the index of the maximum element in $\pmb{y}'_{i, t}$. Then, a OneHot operation is used to get the integer value of a class $Y_{i, t} = \onehot (\pmb{y}_{i, t})$. In this way, gradients can be backpropagated through $\pmb{y}_{i, t}'$. The same approach is also used for all the labels except the last one $Y_H$. Notice that the larger $\tau$, the more uniformly regulated with more stable gradient flow are the sampled values. The typical approach is to decrease $\tau$ as training continues and we can adopt a decrease strategy similar to \cite{jang2016categorical} (Equations \ref{eq:decode_cat}).
\begin{equation}
\begin{split}
    \pmb{y}_{i, t}' &= \tanh \big((\pmb{q}^{i,t}_v + \pmb{g}_v) / \tau \big) \\
    \pmb{y}_{i, t} &= \arg\max \pmb{y}_{i, t}' \\
    Y_{i, t} &= \onehot (\pmb{y}_{i, t})
\end{split}
\label{eq:decode_cat}
\end{equation}

\subsection{Class embedding sub-module}
\label{sec:embedding}
Generated class for each label and time $\hat{Y}_{i, t}$ is a categorical value. Inspired in Word2Vec \cite{rong2014word2vec} in Natural Language Processing, we convert categorical classes to a continuous space to be processed by the following neural networks. 
The embedded continuous value is the conditional signal we need for conditional dependence computing of the next label $Y_{i+1}$. 
To achieve this, an embedding weight table $\pmb{W}_i \in \mathbb{R}^{C_h \times E_i}$ contains all learnable embedding parameters, where $C_h$ is the number of classes in label $Y_i$, $E_i$ is a hyper-parameter of the dimension of continuous space, normally $E_i \ll C_h$. Here, we simply take $E_i = \frac{C_h}{2}$. Each label $Y_i$ has its own embedding table $\pmb{W}_i$. Embedding operation is $\Bar{\pmb{y}}_{i, t} = \pmb{W}_i Y_{i}$, where $\Bar{\pmb{y}}_{i, t}$ is projected continuous vector in a continuous space.

\subsection{Merging sub-module to capture joint conditions}
\label{sec:merging}
The embedded vector $\Bar{\pmb{y}}_{i, t}$ are concatenates with all previous embedded vector $\Bar{\pmb{y}}_{1, t}, \dots, \Bar{\pmb{y}}_{i-1, t}$ and the raw sensors $\pmb{X}_{,t}$. In this way, the next label $Y-i$'s joint conditions of all previous labels and sensors, noted as $p(Y_i | X, Y_1, \dots, Y_{i-1})$ in MLE before, are captured in merged vector as input for the next decoding module $f_{\theta_i}$. The only exception is the last label's embedding vector $\Bar{\pmb{y}}_{H, t}$, which has not concatenation operation, since we have reached the end.

In our proposed Conditional-UNet, a natural question is that what is the best order to sequentially model joint label conditions. This is just another hyper-parameter to be tuned. If there are $H$ labels, there is potentially $(H-1)!$ orders. However, fortunately, there are normally not many labels in real-world applications (e.g. $2$ labels in our head gesture experiment, and there are $2! = 2$ orders to be tuned on).

\section{Experiments}
\label{sec:experiments}
In order to demonstrate and verify the performance of the proposed Conditional-UNet for Co-HAR problem, we conduct experiments as follows: (i) collect a new dataset about head gesture under walk/sit situation through Arduino UNO and other hardwares; (ii) compare our method and its variants with state-of-the-art competing methods; (iii) a qualitative analysis to illustrate effectiveness of our proposed method.

\subsection{Device design and experiment settings}
\label{sec:experiment_setting}
\begin{figure}[!t]
    \centering
    \includegraphics[width=\linewidth, trim={0cm 12cm 3cm 0cm},clip]{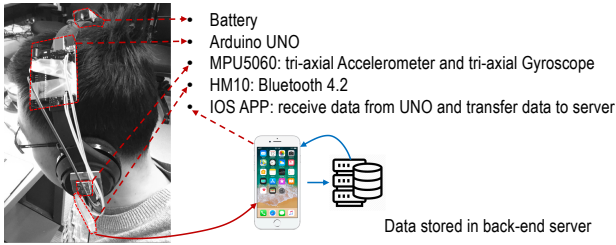}
    \caption{Hardwares and softwares to collect conditional multiple labels and wearable sensor data of tri-axial acceleration and tri-axial gyroscopes through bluetooth communication with an IOS app and backend data storage server.}
    \label{fig:system}
\end{figure}

Hardware design: To our best knowledge, there are no dataset that is collected for Co-HAR yet, especially sensor module locates only at one location of body, so that we can only retrieve mixed signals instead of signals from multiple locations. The types of labels should be conditional dependent and interactively impact each other, so that we need to classify multiple different labels for the same sample. With this in mind, we implement an Arduino UNO module with acceleration and gyroscope sensors located on a headphone (e.g. Figure \ref{fig:system}). The data communication is done through a bluetooth HM-10 module which send data to an IOS iphone app. Also, a camera which is not shown in the figure simultanuously records a user's head gesture and body movement video as the Arduino UNO collect sensor data.

\begin{figure}
    \centering
    \includegraphics[width=\linewidth, trim={0cm 9.5cm 7.5cm 0cm},clip]{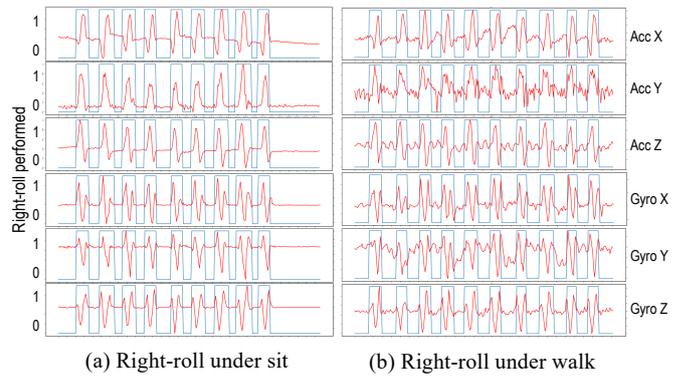}
    \caption{Collected sensor data for right-roll gesture, red lines are sensor signals, blue line indicate if a right-roll is performed (noted $1$) or not (noted $0$): (a) under sit condition. The sensor signals are with clearer patterns; (b) under walk condition. The sensor signals are disturbed from significant body movements.}
    \label{fig:right_roll}
\end{figure}

Briefly, there are the main software component for Arduino UNO. First, it callibrates the device for the initial few seconds. Then, it starts the read of sensor data. Then, it sends the data to a registered bluetooth transmit address to let bluetooth module send data to iphone through an simple IOS app to parse transmitted data and upload to a backend server.
Notice that some basic manual cleaning of the data is done according our recorded videos (around $3-4$ minutes) like removing the starting and ending periods (about a few second duration). We then use synchronized cleaned videos (about $10$ videos for each combination of head gesture classes and walk/sit classes) to manually label head gesture label and walk/sit label. In general, we collect $9$ classes for head gesture label, namely left-roll, right-roll, head-right, head-left, right-lean, left-lean, head-up, head-down, and a null class of no-move (e.g. in Figure \ref{fig:example}), and $2$ classes of walk/sit label, namely if a user is walking or sitting. The baud rate is set as $9600$ bits per second, and our data sample rate is chosen at $\frac{1}{12}Hz$. Under this setting, each ground truth head gestures contains about $20$ samples in a duration of about $1.6$ seconds. The over summary of our collected data is in Table \ref{tab:gesture_stats}. We can see that left roll and right roll takes a lot more time than other head gestures. Also, we can found that duration varies for each head gesture class too. Maximum duration could be $0.3$s more than the minimum duration, which is about $4$ more samples. The visualization of right-roll under both sit and walk condition in Figure \ref{fig:right_roll} also intuitively show such varied duration at different times and also the strong impact from body movement under walk condition. This is an indication that Co-HAR problem is more challenging than just sit without walk.
\begin{table}
\scriptsize
\centering
\caption{Summary of collected head gesture data}
\begin{tabular}{ |m{.5in}|m{.6in}|m{.6in}|m{.75in}| } 
 \hline
 Gesture & class number per sit video & class number per walk video & duration range of a gesture (second) \\ \hline
 head up & 9 & 9 & 1.5 - 1.8\\ \hline
 head down & 9 & 9 & 1.5 - 1.8\\ \hline
 head left & 9 & 10 & 1.5 - 1.8\\ \hline
 head right & 10 & 10 & 1.5 - 1.7\\ \hline
 left lean & 10 & 10 & 1.5 - 1.7\\ \hline
 right lean & 9 & 10 & 1.5 - 1.7\\ \hline
 left roll & 10 & 9 & 1.9 - 2.1 \\ \hline
 right roll & 10 & 9 & 1.9 - 2.1\\ \hline
 no gesture & - & - & - \\ \hline
\end{tabular}
\label{tab:gesture_stats}
\end{table}

\subsection{Competing Methods and Conditional-UNet variants}
Two competing methods including a pre-dated conventional method, and a baseline deep UNet model are used. Two variants of our proposed Conditional-UNet model are also introduced here.

\subsubsection{Surpport Vector Machine (SVM)} SVM is a conventional method widely used pre-dated deep learning methods. We use it as a naive baseline. The six sensor signals are used as raw feature inputs. Two different SVM models are trained separately for head gesture label and walk/sit label.

\subsubsection{UNet baseline (UNet)} This is a baseline multi-label classification based on UNet, a state-of-the-art deep fully convolutional network for HAR \cite{zhang2019human}, which only use one UNet decoding module to output both head gesture label and walk/sit label at the same time without any conditional dependence.

\subsubsection{Dense Head Conditioned on Dense Walk (DHcoDW)} This model is a variant of our Conditional-UNet that first decoding modules model walk/sit label and an encoding module to encode conditional dependence of walk/sit condition, then sequentially, a second decoding module model head label. ``Dense'' means that both labels are classified for each sample (a.k.a. each time step).

\subsubsection{Dense Walk Conditioned on Dense Head (DWcoDH)} This model is another variant of our Conditional-UNet that first decoding modules model head label and an encoding module to encode conditional dependence of head condition, then sequentially, a second decoding module model walk/sit label. Both labels are classified for each sample.


\subsection{Evaluation Metric}
As a classification problem, we use common accuracy score and multi-label F1 score as evaluation metric to compare competing method with our proposed methods. Overall, multi-label F1 score consider both precision and recall in different classes, and is better than accuracy score. We also demonstrate confusion matrix to show the performance of precision and recall for each class of a label.

\subsection{Quantitative Analysis}
\begin{figure*}[!t]
    \centering
    \includegraphics[width=\linewidth, trim={0.5cm 9cm 0cm 0.5cm},clip]{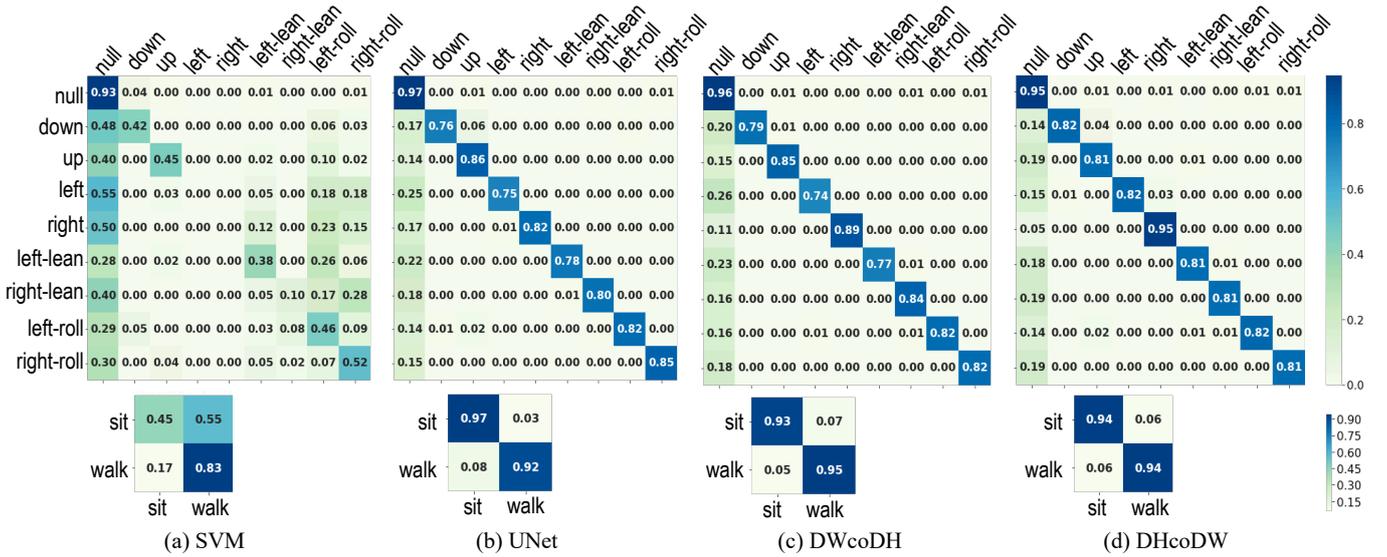}
    \caption{Confusion matrix of multi-label classifications of different methods. Each row is normalized with total sum of each row (number of ground truth classes), and diagonal elements are true positive rate for each class.}
    \label{fig:confusion_matrix}
\end{figure*}
Table \ref{tab:model_performance} contains the experiment results of accuracy and F1 scores by competing methods and variants of our proposed methods. The bolded values are the best one compared to other methods. We can see that SVM fail for head gesture label with a low F1 score of only $35.57\%$, while SVM perform better for walk/sit label with $66.71\%$ F1 score. It indicates that walk/sit has much stronger signals and is easier to be classified. UNet already perform very well as current state-of-the-art method with $90.46\%$ accuracy score and $84.60\%$ F1 score for head gesture label, and $94.94\%$ accuracy score and $94.15\%$ F1 score for walk/sit label. For head gesture label, DHcoDW variant perform about $2\%$ better on accuracy ($92.06\%$) and $3.2\%$ better on F1 ($87.83\%$). The performance on walk/sit label of DHcoDW is only a little worse than UNet ($1.94\%$ less on accuracy and $2.15\%$ less on F1). For DWcoDH variant, head gesture also improve about $1\%$ of accuracy and $2\%$ of F1 score, and walk/sit labels are almost equally as well as UNet baseline. This results show that DHcoDW variant utilize the conditional dependence of walk/sit to better classify head gestures with a little downgrading of walk/sit label. But, DWcoDH's results show that conditional dependence of head gestures promote less performance gain because the stronger signals of walk/sit vanish signals of head movements. This is also a good illustration of advantages of our proposed Conditional-UNet for real-world applications, since head gesture label is more critical and interesting than walk/sit label in real-world applications. Walk/sit condition is more like a noise to be removed than a valuable label to be used.
\begin{table}
\centering
    \captionof{table}{Model performance comparisons}
    \label{tab:model_performance}
    \begin{tabular}{ m{.4in} m{.4in} | m{.4in} m{.4in} m{.4in} m{.4in} m{.4in} }
    \toprule
    Labels & \diagbox[width=.6in]{Metric}{Model} & SVM & UNet & DHcoDW & DWcoDH  \\
    \midrule
    \multirow{2}{*}{Head} 
    & Accuracy & 0.7516 & 0.9046 & \textbf{0.9206} & 0.9128 \\
    & F1       & 0.3557 & 0.8460& \textbf{0.8783} & 0.8638 \\
    \hline
    \multirow{2}{*}{Walk/Sit} 
    & Accuracy & 0.6241 & \textbf{0.9494} & 0.9300 & 0.9426  \\
    & F1       & 0.6671 & \textbf{0.9415} & 0.9201 & 0.9369  \\
    \bottomrule
    \end{tabular}
\end{table}

Confusion matrix of different methods are also shown in Figure \ref{fig:confusion_matrix}, which tell more details about different methods. All confusion matrix values are normalized by total number of ground truth in this class, or in another word by sum of each row (diagonal elements are true positive rate). The most important observation is that two variants of our proposed conditional-UNet achieve significantly gains on head gesture label, because naive UNet model mistakenly classify a large portion of head gesture as null class. Since null class and other head gestures are imbalanced, the accuracy score do not quite reflect the margin of improvement as shown by confusion matrix. By comparing DWcoHD and DHcoDW variants, we can see that DHcoDW achieve a higher true positive rate except head-right, left-lean, left-roll. If we compare the walk/sit label, it is found that DHcoDW variant achive more balance between walk class ($94\%$) and sit class ($93\%$), but both naive UNet and DWcoDH have low performances on walk class. This is another good demonstration that the conditional dependence design help to get more gains by learning challenging body movements during walk condition.


\subsection{Qualitative visualization}
\begin{figure}
    \centering
    \includegraphics[width=\linewidth, trim={0cm 9cm 12cm 0cm},clip]{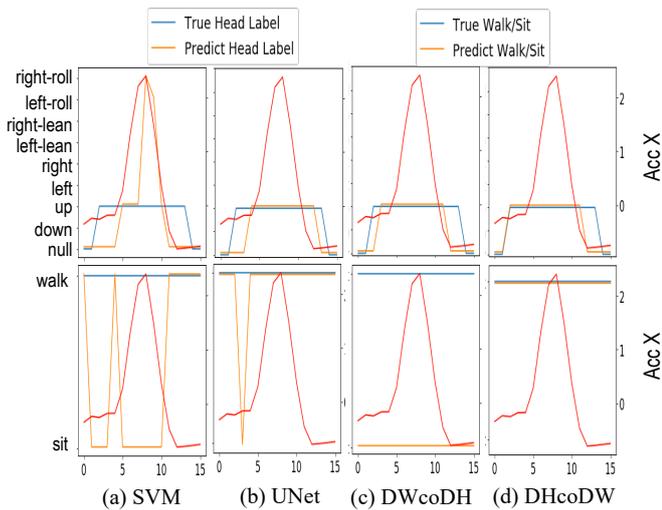}
    \caption{Visualization of results in a selected window which show a raw sensor data (X-axis accelerometer on right Y axis) with classified and ground truth classes (on left Y axis), and X axis is the samples in this window.}
    \label{fig:qualitative_viz}
\end{figure}

We illustrate a few classification results here through visualization of raw sensor, ground truth, and classified classes for both head gesture label and walk/sit label in Figure \ref{fig:qualitative_viz}. Each column is for each methods, the first row is for head gesture label, the second row is for walk/sit label. Left Y axis of the first row plots are different classes of head gestures. Right Y axis of first and second row plots are for raw X-axis accelerometer values. The left Y axis of the second row plots are for walk/sit label. DHcoDW variant of Conditional-UNet outperform a good result (only one sample on the right is classified wrong for head gesture label). DWcoDH unfortunately classify wrong class for walk/sit label, while its classification for head gesture label are very good. In general, this error of one sample ($\frac{1}{12}$ second) could be minor issue for many real-world application.

\section{Conclusion}
We proposed a Coherent Human Activity Recognition problem here with a novel condition-aware deep model of Conditional-UNet to model the joint probability of multiple labels in Co-HAR with explicitly structures to handle conditional dependence in a sequential manner. The conducted experiments show that the proposed method outperforms the pre-dated SVM and state-of-the-art UNet deep model by $3\%$ in F1 score. Moreover, it gets significant gains for different head gestures with a little sacrification of walk/sit label performance. The experiments show that our proposed Conditional-UNet successfully capture the conditional dependence as expected. In this work, a Co-HAR dataset is also contributed to research communities.






\bibliographystyle{IEEEtran}
\bibliography{main.bib}
%



\end{document}